\documentclass[
  ,final            
  ]
  {aipproc}

\layoutstyle{8x11double}

\begin{document}

\title{Super-Massive Neutron Stars}


\author{Paulo C. C. Freire}{
  address={N.A.I.C., Arecibo Observatory, HC3 Box 53995, PR 00612,
  U.S.A.; {\tt pfreire@naic.edu}}
}

\begin{abstract}
We present here the results of Arecibo timing of PSR~B1516+02B,
a 7.95-ms pulsar in a binary system with a $\sim 0.17\,M_{\odot}$
companion and an orbital period of 6.85 days located in the globular
cluster M5. The eccentricity of the orbit ($e = 0.14$) has
allowed a measurement of the rate of advance of periastron:
$\dot{\omega}\,=\,(0.0136\,\pm\,0.0007)^\circ \rm yr^{-1}$.
It is very likely that the periastron advance is due to the effects of
general relativity; the total mass of the binary system is
$(2.14\,\pm\,0.16)\,M_{\odot}$. The small measured mass function
implies, in a statistical sense, that a very large fraction of this
total mass is contained in the pulsar: $M_p \, = \,
(1.94^{+0.17}_{-0.19})\, M_{\odot}$ (1-$\sigma$); there
is a 5\% probability that the mass of this object is below
$1.59\,M_{\odot}$. With the possible exception of PSR~J1748$-$2021B,
this is the largest neutron star mass measured to date. When
combined with similar measurements made previously for
Terzan~5~I and J, we can exclude, in a statistical sense, the ``soft''
equations of state for dense neutron matter, implying that matter at
the center of a neutron star is highly incompressible. There is also
some evidence for a bimodal distribution of MSP masses, the reasons
for that are not clear.
\end{abstract}


\maketitle


\section{Timing of M5B}

Over the past 20 years, more than 130 pulsars have been discovered in
globular clusters (GCs)\footnote{See Scott Ransom's review, in these
Proceedings. For an updated list, see
  \url{http://www2.naic.edu/~pfreire/GCpsr.html}.}. Among the
first discoveries were PSR~B1516+02A and PSR~B1516+02B
\cite{wakp89}. Both of these millisecond pulsars (MSPs) are located in
the GC NGC~5904, also known as M5; for this reason we will refer to
them as M5A and M5B. The latter is a 7.95-ms pulsar in a binary system
with a $\sim 0.17-M_{\odot}$ companion and an orbital
period of 6.86 days. At the time of its discovery, this was the MSP
with the most eccentric orbit known ($e = 0.14$). In the Galaxy, 80\%
of MSPs are found to be in binary systems, and with a single
exception (PSR~J1903+0327, see David Champion's contribution to these
Proceedings) they
are in low-eccentricity orbits with white dwarf (WD) companions. In GCs,
gravitational interactions with neighboring stars can greatly increase
the eccentricity of these binary systems \cite{rh95}; the orbital
eccentricity of M5B is $\sim\, 10^{4}\,-\,10^{5}$ times larger than
that of Galactic MSP-WD systems with similar orbital periods. When
Anderson~et~al.~(1997)\nocite{awkp97} published the timing solutions
of M5A and B, they used the eccentricity of M5B to detect its
periastron advance, but the large relative uncertainty of the
measurement did not allow any astrophysically useful constraints on
the total mass of the binary.

In \cite{fwbh07}, Freire et al. report the results of recent (2001 to
 2007) L-band observations of these two pulsars. The first 2001
observations were part of an Arecibo search for pulsars in GCs, which
found a total of 13 new MSPs \cite{hrs+07}. Three of these were found
in M5, subsequent observations of this GC were made chiefly with
the aim of timing the new discoveries. However, both M5A and M5B are
in the same radio beam as the new pulsars. They are clearly detectable
in the L-band data, allowing for timing of much better (M5A) or
comparable (M5B) quality to that obtained at 430 MHz. Including
those previous data, this provides a much longer total timing baseline
(18 years) and much improved timing parameters.

\subsection{Observations, data reduction and timing}

The L-band observations started in 2001 June. Until 2003, we used
the``old L-Wide'' receiver in the Gregorian
Dome ($T_{sys} = 40$K at 1400 MHz). Since 2003 February, we have been
using the current ``L-Wide'' ($T_{sys} = 25$K at 1400
MHz). The Wide-band Arecibo Pulsar Processors (WAPPs, \cite{dsh00})
made a 3-level digitization of the voltages
of a 100 MHz-wide band for both (linear) polarizations, correlating
them for a total of 256 lags. These were then integrated for a total
of 64$\,\mu$s and the results of both polarizations added in
quadrature and written to disk. At first, only one WAPP was
available. In this case we centered the observing band at
1425~MHz. After 2003, three more WAPPs became available, and we
started observing simultaneously at 1170, 1420 and 1520 MHz, thanks
to the wide frequency coverage of the new L-wide receiver.

The lags were then Fourier transformed to generate power
spectra. These were then dedispersed at the known DM of these pulsars
and folded modulo their spin periods using the {\tt PRESTO} pulsar software
package\footnote{\url{http://www.cv.nrao.edu/~sransom/presto}}.
We then cross-correlated the resulting pulse profiles with the average
pulse profile in the Fourier domain \cite{tay92} to obtain topocentric
times of arrival (TOAs). These were then analyzed with
{\tt
  TEMPO}\footnote{\url{http://www.atnf.csiro.au/research/pulsar/tempo/}},
together with the TOAs derived from the old 430-MHz observations made
from 1989 to 1994. We used the DE~405 Solar System ephemeris
\cite{sta98b} to model the motion of the Arecibo 305-m Radio Telescope
relative to the Solar System Barycenter.

The resulting timing parameters are presented in \cite{fwbh07}.
Astrophysically, the most important new parameter is the rate of
advance of periastron for M5B: 
$\dot{\omega}\,=\,(0.0136\,\pm\,0.0007)^\circ \rm yr^{-1}$.

\subsection{Binary, pulsar and companion masses}
\label{sec:masses}

As argued in \cite{fwbh07}, the $\dot{\omega}$ is solely due to the
effects of general relativity. In this case, we can estimate the total
mass of a binary system:
\begin{equation}
M = \left(\frac{P_b}{2 \pi}\right)^{5/2}
\left[\frac{(1 - e^2)\,\dot{\omega}}{3}\right]^{3/2}
\left(\frac{1}{T_\odot}\right),
\label{eq:totmass}
\end{equation}
where $T_\odot \equiv G M_{\odot} / c^3\,=\,4.925490947 \mu$s. Using
the $\dot{\omega}$ above, we obtain
$M\,=\,(2.14\,\pm\,0.16)\,M_{\odot}$. For the nominal $\dot{\omega}$
and a median inclination of 60$^\circ$, the mass of the companion is
$0.166\,M_{\odot}$ and the mass of the pulsar is
$1.98\,M_{\odot}$. This is well above all the neutron star masses
that have been precisely measured to date.

\begin{figure}
  \label{fig:mass_mass}
  \includegraphics[height=.45\textheight]{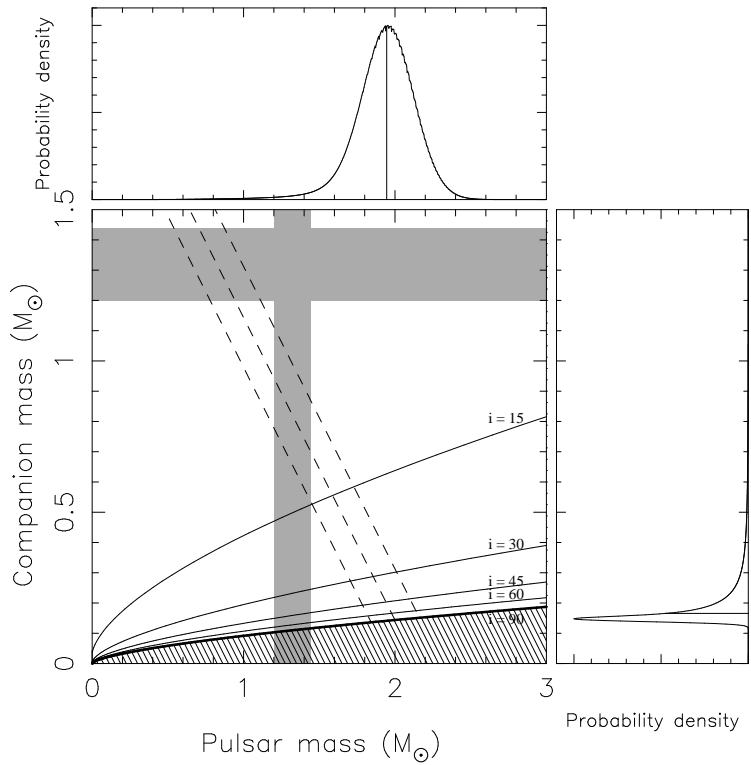}
  \caption{Constraints on the masses of M5B
  and its companion. The hatched region is excluded by knowledge of
  the mass function and by $\sin i \leq 1$.  The diagonal dashed lines
  correspond to a total system mass that causes a general-relativistic
  $\dot{\omega}$ equal or within 1-$\sigma$ of the measured value.
  The five solid curves indicate constant inclinations. The gray bars
  indicate the range of precisely
  measured neutron star masses (from $\sim 1.20\,M_{\odot}$
  to $1.44\,M_{\odot}$). We also display the probability density
  function for the mass of the pulsar ({\em top}) and the mass of the
  companion ({\em right}), and mark the respective medians with
  vertical (horizontal) lines.}
\end{figure}

We calculated a 2-D probability distribution function (pdf) for the
mass of the pulsar and the mass of the companion, assuming that the
pdf for $\dot{\omega}$ is a Gaussian with the half-width equal to
the 1-$\sigma$ uncertainty of the observed $\dot{\omega}$ and an
{\em a priori} constant probability for $\cos i$. The two-dimensional
pdf is then projected in both dimensions, resulting in 1-D pdfs for
the mass of the pulsar and the mass of the companion. These are
displayed graphically in Fig.~\ref{fig:mass_mass}. The pulsar
definitely has a mass smaller than $2.35\,M_{\odot}$, and the
companion has a mass larger than $0.13\,M_{\odot}$, the median and
1-$\sigma$ limits for the pulsar and companion mass are
$1.94^{+0.17}_{-0.19} M_{\odot}$ and $0.164^{+0.10}_{-0.022} M_{\odot}$
respectively. There is a 99\%,
95\% and 90\% probability that the pulsar is more massive than 1.19,
1.59 and $1.69\,M_{\odot}$. There is a 1.3\% probability that $i$ is low
enough to make the neutron star mass fall within the range of NS
masses observed in double neutron star (DNS) systems:
from 1.20~$M_{\odot}$ measured for the companion of PSR~J1756$-$2251
\cite{fkl+05} to 1.44~$M_{\odot}$ measured for PSR~B1913+16
\cite{wt03}.

\begin{table}
\begin{tabular}{ l c r c c c c c c c c}
\hline
  \tablehead{1}{r}{b}{Name PSR}
  & \tablehead{1}{c}{b}{GC}
  & \tablehead{1}{c}{b}{$P$ (ms)}
  & \tablehead{1}{c}{b}{$P_b$ (days)}
  & \tablehead{1}{c}{b}{$e$}
  & \tablehead{1}{c}{b}{$f/M_{\odot}$}
  & \tablehead{1}{c}{b}{$M/M_{\odot}$}
  & \tablehead{1}{c}{b}{$M_c/M_{\odot}$}
  & \tablehead{1}{c}{b}{$M_p/M_{\odot}$}
  & \tablehead{1}{c}{b}{Method}
  & \tablehead{1}{c}{b}{Ref.}
 \\
\hline

J1911$-$5958A & NGC~6752 & 3.26619 & 0.83711 & $<$0.00001 &
0.002688 & 1.58$^{+0.16}_{-0.10}$ & 0.18(2) & $1.40^{+0.16}_{-0.10}$ & Opt. &
\cite{bkkv06} \\

J0024$-$7204H & 47~Tucanae & 3.21034 & 2.35770 & 0.07056 & 0.001927 & 1.61(4) & $> 0.164$ & $< 1.52$ & $\dot{\omega}$ & \cite{fck+03} \\

J1824$-$2452C & M28 & 4.15828 & 8.07781 & 0.84704 & 0.006553 &1.616(7) &
 $> 0.260$ & $< 1.367$ & $\dot{\omega}$ & \cite{brf+07} \\

J1909$-$3744 & - & 2.94711 & 1.53345 & 0.00000 & 0.003122 & 1.642(24) & 0.2038(22) &
1.438(24) & $r,s$ & \cite{jhb+05} \\

\hline

B1516+02B & M5 & 7.94694 & 6.85845 & 0.13784 & 0.000647 & 2.14(16) &
$> 0.13$ & $< 2.35$  & $\dot{\omega}$ & \cite{fwbh07} \\

J1748$-$2446I & Terzan~5 & 9.57019 & 1.328 & 0.428 & 0.003658 & 2.17(2) &
$> 0.24$ & $< 1.96$  & $\dot{\omega}$ & \cite{rhs+05} \\

J1748$-$2446J & Terzan~5 & 80.3379 & 1.102 & 0.350 & 0.013066 &
 2.20(4) &
$> 0.38$ & $< 1.96$  & $\dot{\omega}$ & \cite{rhs+05} \\

J0514$-$4002A & NGC~1851 & 4.99058 & 18.7852 &
0.88798 & 0.145495 & 2.453(14) &
$> 0.96$ & $< 1.52$ & $\dot{\omega}$ & \cite{frg07} \\

J1748$-$2021B & NGC~6440 & 16.76013 & 20.5500 & 0.57016 & 0.000227 & 2.91(25) &$> 0.11$ & $< 3.3$ & $\dot{\omega}$ & \cite{frb+07} \\

\hline

\end{tabular}
\caption{
Millisecond pulsar masses. Terzan 5 J is not technically a MSP, its spin
period is longer than what we find in some DNS systems. However,
given the similarity of its orbital parameters to those of
Terzan~5~I, we assume that it had a similar formation history.
Due to their high eccentricities, M28C and NGC~1851A were likely
formed in an exchange interaction.
}
\label{tab:msp_masses}

\end{table}

\section{Statistical evaluation of mass measurements}

A list of mass estimates for millisecond pulsars is presented in
Table~\ref{tab:msp_masses}. Most of these were derived for MSPs in
GCs. The reason is, of course, than in GCs the MSP-WD orbits can be
perturbed, resulting in much higher eccentricities than possible in
the Galaxy. These eccentricities allow the measurement of
post-Keplerian (PK) effects like $\dot{\omega}$. Eventually other PK
effects will be measurable, like the Einstein delay
($\gamma$). However, GC pulsars are generally rather faint, therefore
such measurements will require some time.

For that reason, and as in the case of M5B, most of the estimates in
Table \ref{tab:msp_masses} are based on measurements of $\dot{\omega}$
only. These are ``incomplete'' measurements, in the sense
that one more PK parameter is necessary to have an
unambiguous determination of the pulsar mass. Nevertheless, unambiguous
upper limits for the mass of the pulsar and lower limits for the mass
of the companion can always be obtained in these cases. Furthermore,
in systems where the mass function is small (as for M5B) there is a much
greater probability of most of the mass of the binary belonging to the
pulsar itself (see above).

From Table~\ref{tab:msp_masses}, we can see that the assumption
that the $\dot{\omega}$ is relativistic yields for M5B the largest
neutron star mass presently known, with the possible exception of
PSR~J1748$-$2021B (NGC~6440B, see Paulo Freire's contribution on the
new pulsars in NGC~6440 and NGC~6441 in these Proceedings).

\subsection{Statistical evidence for high neutron star masses}

One of the interesting features of Table~\ref{tab:msp_masses} is that
as the total binary mass increases, the mass function does not
increase, with a single exception: PSR~J0514$-$4002A (see Paulo
Freire's contribution on NGC 1851A in these Proceedings). This system has
a highly eccentric orbit that is thought to have resulted from an exchange
interaction \cite{fgri04}. If the increase in total mass for the other
systems was due solely to an increase in the mass of the companion, then
there should be a more general trend to higher mass functions, i.e.,
PSR~J0514$-$4002A should be the rule among the massive binaries, not
the exception.

It could happen that the separation between the ``light'' and
``massive'' binaries (shown by the horizontal line in
Table~\ref{tab:msp_masses}) is due merely to a significant classical
contribution to $\dot{\omega}$  in the latter group. This could
explain {\em all} the anomalously high masses produced by the
$\dot{\omega}$ estimates. However, it is unlikely that the extra
classical contributions have exactly the values required to
make the mass estimates nearly identical to each other. Furthermore,
a possible cause of an extra contribution to
$\dot{\omega}$ is tidal effects due to an extended companion. If the
presence of extended companions were to explain the large measured values
of $\dot{\omega}$, then it should also lead on average to tidal
circularization of the orbits of the ``massive'' binaries, particularly
for those with shorter orbital periods (even if that
is not guaranteed in individual cases, like M5B, which could have been
perturbed recently). In reality, the ``massive''
binaries are not less eccentric than the ``light'' binaries.

\subsection{Statistical combination of mass pdfs}
\label{sec:massives}

\begin{figure}
  \label{fig:psr_masses}
  \includegraphics[height=.44\textheight]{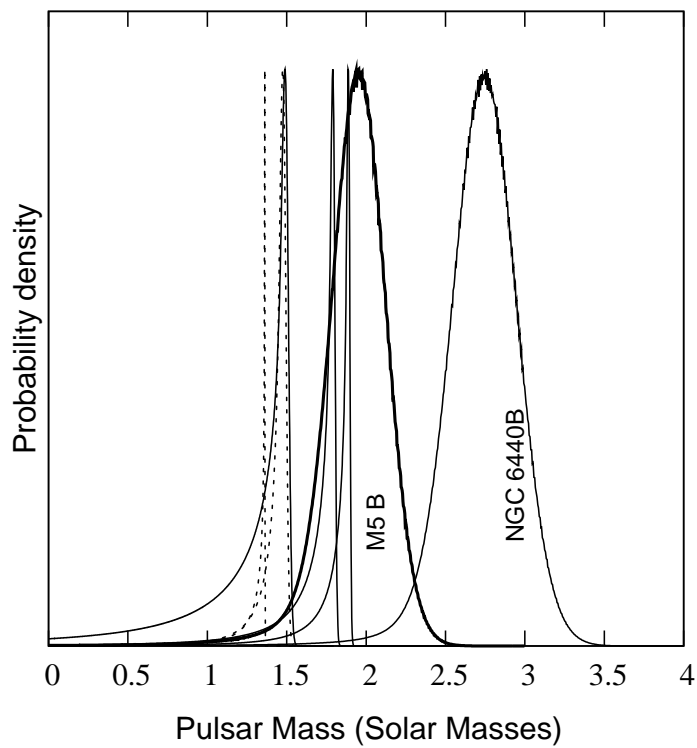}
  \caption{Probability distribution functions (pdfs) for seven of the
  MSPs in Table \ref{tab:msp_masses}. The mass pdfs of the MSPs in the least
  massive binaries (those with $M < 2 M_{\odot}$) are represented by
  the dashed curves. Despite the limitation of being calculated solely
  from $\dot{\omega}$, the pdfs capture well the peak in NS mass that
  is known to occur at $1.2 - 1.4\,M_{\odot}$, suggesting that the
  peak at $1.8 - 2.0\,M_{\odot}$ is a also a real feature.}
\end{figure}

The mass pdfs for the 7 MSPs in Table~\ref{tab:msp_masses} with
$\dot{\omega}$ measurements were calculated as described
above for M5B, and displayed graphically in Fig.~\ref{fig:psr_masses}.
The mass medians for Ter~5~I, J and NGC~6440B are 1.87, 1.76
\cite{rhs+05} and 2.73 \cite{frb+07} respectively. All these
pdfs have long, low-probability tails towards the low masses,
corresponding to improbably low orbital inclinations.

Combining the pdfs for the masses of Ter~5~I and J, \cite{rhs+05}
reached the conclusion that at least one of these is more
massive than 1.48 and $1.68\,M_{\odot}$ with 99\% and 95\%
confidence levels. Combining those pdfs with the mass pdf for M5B, the
probability that all of these pulsars are less massive than 1.72
and $1.79\,M_{\odot}$ is only 1 and 5\% respectively. These limits
introduce some of
the most stringent constraints to date on the bulk behavior of cold,
super-dense matter. They exclude many of the ``soft'' equations of
state that have been proposed to model that behavior \cite{lp07}.

From NGC~6440B alone, we can derive a 99\% probability for that pulsar
having a mass larger than $2.01\,M_{\odot}$, which, if confirmed, would be
an even tighter constraint to the EOS. If all $\dot{\omega}$s are
relativisitc, the probability of all these massive binaries having
``normal'' ($1.2\,<\,M_p\,<\,1.44 M_{\odot}$) masses is $9 \times 10^{-9}$.

\section{Formation}

The mass pdfs in Fig.~\ref{fig:psr_masses} suggest that the
distribution of MSP masses is bi-modal, with NGC~6440B as a
super-massive outlier. PSR~J0514$-$4002A and the MSPs in the ``light''
($M\,<\,2\,M_{\odot}$) binaries have masses smaller than
$1.5\,M_{\odot}$, i.e., they are not significantly more massive than
mildly recycled neutron stars, despite having spin frequencies of
hundreds of Hz. In particular, the case of M28~C shows that MSPs can
be recycled by accreting $<\,0.15\,M_{\odot}$ from their
companions. Other MSPs are significantly more massive, it is not clear
why they are so.

It could happen that they were born that way. A bimodal (or tri-modal)
distribution like that of Fig.~\ref{fig:psr_masses} is exactly what is
predicted by hydrodynamical core collapse simulations \cite{tww96}:
stars below $\sim 18\,M_{\odot}$ are expected to form $\sim
1.20-1.35\,M_{\odot}$ NSs. Stars with masses between $18 - 20\, M_{\odot}$
form $1.8\,M_{\odot}$ NSs. Above $20\,M_{\odot}$, stars experience
partial fall-back of material that can significantly increase the mass
of the remnant, making it either a super-massive NS or, if its mass
is above the maximum stable neutron star mass, a black hole.

This possibility raises the question of why such massive NSs
have never been found in the 9 known DNS systems (which have a total
of 18 NSs). Given the narrow range of progenitor masses ($18 - 20\,
M_{\odot}$), it is possible that massive NSs are relatively rare. However,
50\% of the NSs in eccentric binary MSPs in Table~\ref{tab:msp_masses}
are massive.

It is possible that all these NSs started instead with similar
masses. In the case of MSPs, the accretion episode is much longer than
for the recycled pulsars in DNS systems, with a potentially (but not
necessarily) larger mass transfer. This is a natural explanation for
why we only see massive NSs as MSPs but not in DNS systems. If this
was the case, we should then expect that the more massive MSPs, having
accreted more mass and angular momentum, should spin faster than the
less massive MSPs. Table \ref{tab:msp_masses} shows that the opposite
is true: the more massive MSPs spin more {\em slowly} ($\nu
\,<\, 125\,$Hz) than the less massive MSPs ($\nu\,>\,200\,$Hz). More
statistics are needed to verify the significance of this relation; but
if it holds, then there might be two MSP recycling
mechanisms, one of them transmitting more angular momentum and the
other more mass. If, on the other hand, neutron stars start with
different masses, and the more massive NSs have a higher moment of
inertia, then transferring the same amount of angular momentum to a
massive NS will cause a smaller increase in the spin frequency.


\begin{theacknowledgments}
The Arecibo Observatory, a facility of the National Astronomy and
Ionosphere Center, is operated by Cornell University under a
cooperative agreement with the National Science Foundation. I thank my
collaborators in the GC pulsar searches over the years: many were
students, like Steve B\'egin, Ryan Lynch, Jeniffer Katz, Lucy Frey
and Ben Sulman, others were a bit more experienced, like
Andrew G. Lyne, Richard N. Manchester, Vicky Kaspi, Alex
Wolszczan and Yashwant Gupta. Special thanks go to Fernando Camilo,
Michael Kramer, Duncan Lorimer, Ingrid Stairs, Jason Hessels
and, of course, Scott Ransom.
\end{theacknowledgments}



%
%

\end{document}